\documentclass[aps,prd,preprint,groupedaddress,showpacs,preprintnumbers]{revtex4-1}
\usepackage{amssymb,amsmath}
\usepackage[dvips]{graphicx}
\usepackage{color}

\usepackage{ulem}

\begin{document}

\preprint{KEK-TH-1698}

\title{Momentum spectra of particles produced in a single pulse of an electric field}

\author{Takashi Arai}
\email[]{arai-t@sys.i.kyoto-u.ac.jp}

\affiliation{Graduate School of Informatics, Kyoto University, Kyoto 606-8501, Japan}


\begin{abstract}
We study particle creation in a single pulse of an electric field in scalar quantum electrodynamics.
We first identify parameter regions of the theory where the dynamical pair creation and Schwinger mechanism respectively dominate each other.
Then, analytical expressions for the total characteristics of particle creation are determined for the case where the Schwinger mechanism dominates.
We also compare our results with those produced in a constant electric field with a finite-time interval.
These results coincide at a strong field regime, however they differ in general field strength.
We identify the reason of this difference with a nonperturbative effect of high-frequency photons in external electric fields.
\end{abstract}


\maketitle

\section{Introduction}
Particle creation from a vacuum by an external electric field is a common phenomenon in quantum field theory.
This phenomenon was first studied by Schwinger in a spatially homogeneous constant electric field~\cite{Schwinger}, which is now known as the Schwinger mechanism, and then the study was extended to electric fields with various time-dependences.
For example, particle creation in a spatially homogeneous single pulse of an electric field has been studied~\cite{Nikishov, Nikishov_pulse}, and the methods of treating particle creation in an arbitrary time-dependent electric field have been developed~\cite{Itzykson, Grib, Gitman}.
Since the single pulse of an electric field is an idealized form of an electric field realized by two colliding laser beams, particle creation in an alternating electric field has been studied for a more realistic situation~\cite{Nikishov2, Mostepanenko}.
However, it was found that the creation rate of the Schwinger mechanism is exponentially suppressed by the mass of the produced particle and an extremely strong electric field is necessary for the actual observation of this phenomenon.
Thus, this phenomenon has never been observed experimentally.
Therefore, it is still unclear to what extent theoretical prediction actually captures the physics of particle creation.

However, the study of particle creation in electric fields has now attracted renewed attention because of the recent development of the strong laser technique, in which the electric field strength nears the critical value of the Schwinger mechanism.
Recently, it was found that the critical threshold of electric fields could be lowered by the superposition of two pulsed electric fields with different frequencies.
This is called the dynamically assisted Schwinger mechanism with which particle creation can be observed in the electric field below the critical strength~\cite{Dunne, Alkofer, Schutzhold}.
Furthermore, it is indicated that the Schwinger mechanism is testable indirectly in the condensed matter system of a graphene single monolayer~\cite{Allor, Beneventano, Mostepanenko_graphene}, in which the electrons inside are described approximately by the massless pseudo-relativistic Dirac equation.

In this paper, we study a formal aspect of particle creation in a spatially homogeneous time-dependent electric field.
Generally in a time-dependent electric field, a difference in the constant value of the vector potential develops between in and out asymptotic regions, even though the electric field asymptotically vanishes.
This constant term affects the dispersion relation of the asymptotic mode functions.
In other words, the canonical momentum of the asymptotic mode functions does not coincide with the kinetic momentum due to the existence of the constant term of the vector potential.
This fact complicates the interpretation of the obtained results.
In fact, in a preceding study, the canonical momentum is confused with the momentum a particle possesses because of its motion~\cite{Mostepanenko_graphene}.
Moreover in our view, the discrepancy of the two momenta leads not only to inappropriate interpretation, but also to the incorrect result that the momentum spectra produced by a pulsed electric field and a constant electric field coincide when we take a limit of an infinite-time interval~\cite{Nikishov_pulse, Nikishov_Nucl}.

In this paper we first define asymptotic particles in which the canonical momentum is identical to the kinetic ones.
Then we investigate parameter regions of the theory where the dynamical pair creation and Schwinger mechanism respectively dominate each other.
We revisit the total characteristics such as the total number of produced particles and the vacuum-to-vacuum transition amplitude for the case where the Schwinger mechanism dominates.
Furthermore, we compare our results with those in a T-constant electric field (a constant field with a finite-time interval~\cite{Gitman7162}).
As a result, it is found that in a strong field regime a T-constant field and pulsed electric field produce quantitatively the same particle spectrum even at a finite-time interval, however in general field strength these results differ.

For a realistic situation related to actual experimental observations, one has to consider a fermionic field.
However, it is sufficient to consider a scalar field for our purpose of seeing how our particle definition works and comparing results with T-constant fields.
Therefore, in this paper, we consider scalar quantum electrodynamics in a single pulse of an electric field to circumvent technical complexities.
We use the Bogoliubov transformation between the in and out asymptotic particles to treat particle creation by an external field.
The quantum effect of electromagnetic interaction is ignored in a similar way as in most preceding studies.

This paper is organized as follows.
In Sec.~\ref{sec:Bogoliubov}, we review the mode expansions of the quantum field and derive the Bogoliubov transformation for asymptotic particles.
In Sec.~\ref{sec:Momentum_spectrum}, we consider the specific cases in which the Schwinger mechanism and the dynamical pair creation, respectively, dominate each other.
The parameter regions of the theory are identified in each case.
In Sec.~\ref{sec:Implication}, analytic expressions for the total number of produced particles and the vacuum-to-vacuum transition amplitude are derived for the case in which the Schwinger mechanism dominates.
Furthermore, we compare our results with those produced in a T-constant electric field.
Sec.~\ref{sec:Conclusion} is devoted to conclusions.
In this paper, we use the unit system of $\hbar = c =1$ and the signature of the metric $(+ - - - )$.
The charge of the particles is $-e$, a negative quantity.

\section{Bogoliubov transformation between asymptotic particles \label{sec:Bogoliubov}}
In this paper, we consider a complex scalar field theory with the Lagrangian given by
\begin{equation}
S[\phi]=\int d^4 x [\eta^{\mu\nu} \nabla_{\mu}\phi^{\ast} \nabla_{\nu}\phi -m^2 \phi^{\ast}\phi ],
\end{equation}
where $\nabla_{\mu}$ is the covariant derivative, $\nabla_{\mu}\phi=(\partial_{\mu}-i e A_{\mu})\phi$, and $A_{\mu}$ is an external gauge field.
The equation of motion for the scalar field is given by
\begin{equation}
[\partial^2-i e \partial_{\mu} A^{\mu}-2 i e A^{\mu}\partial_{\mu}-e^2 (A_{\mu})^2+m^2] \phi(x)=0.
\end{equation}
Now, we consider the behavior of the scalar field in a spatially homogeneous single pulse of an external electric field, given by $E_3(t)=E \cosh^{-2} \Omega t$.
This electric field is realized by the following vector potential,
\begin{gather}
A_1=A_2=0, \\
A_3=\frac{E}{\Omega} (1+\tanh \Omega t),
\end{gather}
where we have taken the gauge condition as $A_0=0$ and set $\lim_{t \rightarrow -\infty}A_3(t)=0$ without loss of generality.
In this gauge potential, the equation of motion reads as
\begin{equation}
[\partial_t^2-\partial_i^2+2 i e A_3 \partial_3+e^2 (A_3)^2+m^2]\phi(x)=0.
\end{equation}
Using the Fourier transformation in the spatial direction $\phi(x)=\int \frac{d^3 p}{(2\pi)^3} \phi(p)e^{i \mathbf{p \cdot x}}$, one can transform the equation of motion to
\begin{equation}
\Bigl[ \partial_t^2+\mathbf{p}^2-2 p_3 \frac{e E}{\Omega}(1+\tanh \Omega t)+ \frac{(eE)^2}{\Omega^2} (1+\tanh \Omega t)^2+m^2 \Bigr]\phi(p)=0.
\end{equation}
This equation can be solved as a hypergeometric differential equation~\cite{Nikishov}.
In fact, the change of the time variable $u=\frac{1}{2}(1+\tanh \Omega t)$ transforms the equation as
\begin{equation}
\begin{split}
 \biggl\{ &u^2 (1-u)^2 \frac{d^2}{du^2} +u (1-u) (1-2 u) \frac{d}{du} \\
& +\frac{1}{4}\frac{\mathbf{p}^2+m^2}{\Omega^2}
-\frac{2}{4}\frac{p_3}{\Omega} \frac{2eE}{\Omega^2}u  +\frac{1}{4}  \Bigl(\frac{2eE}{\Omega^2}\Bigr)^2 u^2 \biggr\} \phi(p)=0.
\end{split}
\end{equation}
Furthermore, if we set
\begin{equation}
\phi(p)=u^a (1-u)^b f(u),
\end{equation}
where
\begin{gather}
a=-\frac{i}{2}\frac{\omega}{\Omega}, \\
b=\frac{i}{2}\frac{\omega^+}{\Omega}, \\
\omega=[m^2+\mathbf{p}^2]^{\frac{1}{2}}, \\
\omega^{\pm}=\Bigl[ m^2+p_{\perp}^2+(p_3\mp \frac{2eE}{\Omega})^2 \Bigr]^{\frac{1}{2}},
\end{gather}
and $p_{\perp}$ is a momentum perpendicular to the applied electric field, $p_{\perp}^2=p_1^2+p_2^2$, the function $f(u)$ satisfies the following differential equation,
\begin{equation}
\begin{split}
\biggl\{ u(1-u) \frac{d^2}{du^2} &+ [1+2a -u (2+2a+2b)] \frac{d}{du} \\
&-(a+b+2ab)+\frac{1}{4} \Bigl[ \frac{2 (\mathbf{p}^2+m^2)}{\Omega^2}-\frac{2 p_3}{\Omega}\frac{2eE}{\Omega^2} \Bigr] \biggr\} f(u)=0.
\end{split}
\end{equation}
This is precisely the form of the hypergeometric differential equation
\begin{equation}
\Bigl[ z(1-z) \frac{d^2}{dz^2}+[\gamma-(\alpha+\beta+1) z ]\frac{d}{dz}-\alpha \beta\Bigr] f(z)=0,
\end{equation}
with
\begin{gather}
\alpha=\frac{1}{2}+a+b+\frac{i}{2}c, \\
\beta=\frac{1}{2}+a+b-\frac{i}{2}c, \\
\gamma=1+2a, \\
c=\Bigl[ \bigl(\frac{2eE}{\Omega^2}\bigr)^2-1\Bigr]^{\frac{1}{2}}.
\end{gather}
Thus, the scalar field can be expanded in terms of independent solutions of the hypergeometric differential equation.

Here, we define the asymptotic particles using the asymptotic form for the mode expansions at $t \rightarrow \pm \infty$.
First, in the case of the asymptotic in region $t \rightarrow -\infty$, we use the two independent solutions that are regular at $u=0$.
Then, the scalar field is expanded as
\begin{equation}
\begin{split}
\phi(x) =&\int \frac{d^3 p}{(2\pi)^3} \frac{1}{\sqrt{2 \omega}} \biggl\{
a_p \bigl[ u^a (1-u)^b {}_2F_1(\alpha,\beta,\gamma;u) \bigr] (p) \\
&+ b_{-p}^{\dagger} \bigl[ u^{-a}(1-u)^b {}_2F_1(1+\alpha-\gamma,1+\beta-\gamma,2-\gamma;u) \bigr] (p) \biggr\} e^{i \mathbf{p \cdot x}},
\end{split}
\end{equation}
where $[\cdots](p)$ denotes that $a$ and $b$ in $[\cdots]$ are evaluated about the momentum $p$.
The mode expansion has the following asymptotic form at $t\rightarrow -\infty$,
\begin{equation}
\phi(x) \sim \int \frac{d^3p}{(2\pi)^3}\frac{1}{\sqrt{2\omega}}\bigl[ a_p e^{-i \omega t+i{\mathbf{p \cdot x}}}+b_p^{\dagger} 
e^{i \omega t-i \mathbf{p \cdot x}} \bigr],
\label{eq:mode_expansion_free}
\end{equation}
which coincides with that of the free field.
Therefore, when we quantize the scalar field imposing the canonical commutation relation $[ \phi(x), \pi(x')]=i \delta^3 (\mathbf{x-x'})$, where $\pi(x)$ is the canonical momentum variable conjugate to $\phi(x)$, the coefficients $a_p$ and $b_p^{\dagger}$ are interpreted as the creation and annihilation operators for the particles and antiparticles that satisfy $[a_p, a_{p'}^{\dagger} ]=[b_p, b_{p'}^{\dagger} ]=(2\pi)^3 \delta^3 (\mathbf{p-p'})$.
We can define the vacuum state at $t \rightarrow -\infty$ by the condition $a_p |0\rangle_{\mathrm{in}} =b_p|0\rangle_{\mathrm{in}}=0$.

Next, in the case of the asymptotic out region $t\rightarrow \infty$, we expand the field in terms of two independent solutions that are regular at $u=1$,
\begin{equation}
\begin{split}
\phi(x) =&\int \frac{d^3 p}{(2\pi)^3} \frac{1}{\sqrt{2 \omega^+}} \biggl\{
\tilde{c}_p \bigl[ u^a (1-u)^b {}_2F_1(\alpha,\beta,1+\alpha+\beta-\gamma;1-u)\bigr] (p) \\
&+\tilde{d}_{-p}^{\dagger} \bigl[ u^{a}(1-u)^{-b} {}_2F_1(\gamma-\alpha,\gamma-\beta,1+\gamma-\alpha-\beta;1-u) \bigr] (p) \biggr\}
e^{i \mathbf{p \cdot x}}.
\end{split}
\label{eq:expansion_inappropriate}
\end{equation}
The mode expansion has the following asymptotic form at $t\rightarrow \infty$,
\begin{equation}
\phi(x) \sim \int \frac{d^3p}{(2\pi)^3} \frac{1}{\sqrt{2\omega^+}} \bigl[\tilde{c}_p e^{-i \omega^+  t}
+\tilde{d}_{-p}^{\dagger} e^{i \omega^+ t} \bigr] e^{i \mathbf{p \cdot x}}.
\end{equation}
Then, it is tempting to interpret the coefficients $\tilde{c}_p$ and $\tilde{d}_{-p}^{\dagger}$ as the creation and annihilation operators at the out region in a similar way.
However, this interpretation causes a difficulty.
In fact, the asymptotic mode functions have a dispersion relation different to that of the free field $\omega^+ =[m^2+p_{\perp}^2+(p_3-\frac{2eE}{\Omega})^2 ]^{\frac{1}{2}}$.
In a preceding study, this fact is represented as the discrepancy between the canonical momentum, which is the momentum of the Fourier expansion of the field $e^{i \mathbf{p \cdot x}}$ and the kinetic momentum that a particle possesses because of its motion.
This discrepancy derives from the constant term in the vector potential.
In fact, if we consider a theory with a constant vector potential $A_3=a_3$, the equation of motion reads as
\begin{equation}
\begin{split}
&\bigl[\partial_t^2-\partial_i^2+2 i e a_3 \partial_3+e^2 (a_3)^2+m^2 \bigr]\phi(x), \\
=&\bigl[ \partial_t^2-\partial_{\perp}^2-(\partial_3-i e a_3)^2+m^2 \bigr]\phi(x)=0,
\end{split}
\label{eq:eom}
\end{equation}
where we set $\partial_{\perp}^2=\partial_1^2+\partial_2^2$.
In this way, the constant term in the vector potential affects the dispersion relation of the mode functions.

In general, when we consider a time-dependent electric field, a difference in the constant term of the vector potential necessarily develops between the in and out asymptotic regions, even though the electric field asymptotically vanishes.
Therefore, one cannot set $A_3=0$ at both the in and out asymptotic regions simultaneously using the gauge degree of freedom of $A_3$.
In our case, for example, the gauge degree of freedom of $A_3$ is already used to set $A_3=0$ at $t \rightarrow -\infty$; thus, we cannot set $A_3=0$ at $t \rightarrow \infty$.
However, one can think of this term as the degree that does not contribute to the physics, since the constant term of the vector potential is intrinsically the gauge degree of freedom.

Therefore, in this paper, we propose the particle definition in which the constant term is isolated as a gauge phase.
In fact, one can transform Eq.~(\ref{eq:eom}) to that of the free field by isolating the gauge constant as a phase factor $\phi(x)=e^{i e a_3 x_3} \varphi(x)$,
\begin{equation}
e^{i e a_3 x_3} \bigl[\partial_t^2-\partial_i^2+m^2 \bigr]\varphi(x)=0.
\end{equation}
By the same prescription to Eq.~(\ref{eq:expansion_inappropriate}), the mode expansion at $t \rightarrow \infty$ is given as follows:
\begin{equation}
\begin{split}
\phi(x) =&\int \frac{d^3 p}{(2\pi)^3} \frac{1}{\sqrt{2 \omega}} \biggl\{
c_p \bigl[ u^a (1-u)^b {}_2F_1(\alpha,\beta,1+\alpha+\beta-\gamma;1-u) \bigr] (p+\tfrac{2eE}{\Omega}) \\
&+
d_{-p}^{\dagger} \bigl[ u^{a}(1-u)^{-b} {}_2F_1(\gamma-\alpha,\gamma-\beta,1+\gamma-\alpha-\beta;1-u) \bigr] (p+\tfrac{2eE}{\Omega})
 \biggr\} e^{i \mathbf{p \cdot x}} e^{i \frac{2eE}{\Omega} x_3}.
\end{split}
\label{eq:mode_expansion_out}
\end{equation}
This has the same asymptotic form at $t\rightarrow \infty$ as that of the free field, Eq.~(\ref{eq:mode_expansion_free}).
Thus, we can interpret the coefficients of the mode expansion of the field $\varphi(x)$, $c_p$ and $d_p^{\dagger}$, as the creation and annihilation operators of the particle and antiparticle at the asymptotically out region.
That is, when one imposes the canonical commutation relation $[\phi(x),\pi(x')]=i \delta^3 (\mathbf{x-x'})$, the ladder operators satisfy $[c_p, c_{p'}^{\dagger}]=[d_p, d_{p'}^{\dagger}]=(2\pi)^3 \delta^3 (\mathbf{p-p'})$.
We define the vacuum state at the asymptotically out region as $c_p|0\rangle_{\mathrm{out}}=d_p |0\rangle_{\mathrm{out}}=0$.

These ladder operators at the asymptotic in and out regions are related to each other by the Bogoliubov transformation.
In fact, the identities of the hypergeometric function 
\begin{equation}
\begin{split}
{}_2F_1(\alpha,\beta,\gamma;z)=&\frac{\Gamma(\gamma)\Gamma(\gamma-\alpha-\beta)}
{\Gamma(\gamma-\alpha)\Gamma(\gamma-\beta)}
{}_2F_1(\alpha,\beta,\alpha+\beta+1-\gamma;1-z) \\
&+\frac{\Gamma(\gamma)\Gamma(\alpha+\beta-\gamma)}
{\Gamma(\alpha)\Gamma(\beta)}
(1-z)^{\gamma-\alpha-\beta} {}_2F_1(\gamma-\alpha, \gamma-\beta,1+\gamma-\alpha-\beta;1-z),
\end{split}
\end{equation}
\begin{equation}
{}_2F_1(\alpha,\beta,\gamma;z)=(1-z)^{\gamma-\alpha-\beta} 
{}_2F_1(\gamma-\alpha,\gamma-\beta,\gamma;z),
\end{equation}
enable us to transform the mode expansion at the in region in the following way:
\begin{equation}
\begin{split}
\phi(x) =\int \frac{d^3 p}{(2\pi)^3} \frac{1}{\sqrt{2 \omega}} \Biggl\{&
\biggl[ a_p \frac{\Gamma(\gamma)\Gamma(\gamma-\alpha-\beta)}{\Gamma(\gamma-\alpha)\Gamma(\gamma-\beta)}
+b_{-p}^{\dagger} \frac{\Gamma(2-\gamma)\Gamma(\gamma-\alpha-\beta)}{\Gamma(1-\alpha)\Gamma(1-\beta)} \biggr] (p) \\
& \bigl[ u^a (1-u)^b {}_2F_1(\alpha,\beta,\alpha+\beta+1-\gamma;1-u) \bigr] (p) \\
+&
\biggl[ a_{p}\frac{\Gamma(\gamma)\Gamma(\alpha+\beta-\gamma)}{\Gamma(\alpha)\Gamma(\beta)}
+b_{-p}^{\dagger}\frac{\Gamma(2-\gamma)\Gamma(\alpha+\beta-\gamma)}{\Gamma(1+\alpha-\gamma)\Gamma(1+\beta-\gamma)} \biggr] (p) \\
& \bigl[ u^a (1-u)^{-b} {}_2F_1(\gamma-\alpha,\gamma-\beta,1+\gamma-\alpha-\beta;1-u) \bigr](p) \Biggr\} e^{i \mathbf{p \cdot x}}.
\end{split}
\end{equation}
Shifting the momentum variable as $p_3 \rightarrow p_3 +\frac{2eE}{\Omega}$ and comparing the expression to the mode expansion at the out region Eq.~(\ref{eq:mode_expansion_out}), one can read the Bogoliubov transformation as follows:
\begin{equation}
\begin{split}
c_p=&\sqrt{\frac{\omega}{\omega^-}}
\left[ a_{p+\tfrac{2eE}{\Omega}} \frac{\Gamma(\gamma)\Gamma(\gamma-\alpha-\beta)}{\Gamma(\gamma-\alpha)\Gamma(\gamma-\beta)}
+b_{-(p+\tfrac{2eE}{\Omega})}^{\dagger} \frac{\Gamma(2-\gamma)\Gamma(\gamma-\alpha-\beta)}{\Gamma(1-\alpha)\Gamma(1-\beta)} \right](p+\tfrac{2eE}{\Omega}), \\
d_{-p}^{\dagger}=&\sqrt{\frac{\omega}{\omega^-}}
\left[ a_{p+\tfrac{2eE}{\Omega}}\frac{\Gamma(\gamma)\Gamma(\alpha+\beta-\gamma)}{\Gamma(\alpha)\Gamma(\beta)}
+b_{-(p+\tfrac{2eE}{\Omega})}^{\dagger}\frac{\Gamma(2-\gamma)\Gamma(\alpha+\beta-\gamma)}{\Gamma(1+\alpha-\gamma)\Gamma(1+\beta-\gamma)} \right]
(p+\tfrac{2eE}{\Omega}),
\end{split}
\end{equation}
where the shift of the momentum in the operator indices $a_{p+2eE/\Omega}$ is understood to be taken only in the direction of the applied electric field.
If we express the Bogoliubov coefficients by $\alpha_p$ and $\beta_p$ as
\begin{equation}
\begin{split}
&c_p=\alpha_p a_{p+e a_3}+\beta_p^{\ast} b_{-(p+e a_3)}^{\dagger}, \\
&d_{-p}^{\dagger}=\beta_p a_{p+e a_3}+\alpha_p^{\ast} b_{-(p+e a_3)}^{\dagger},
\end{split}
\end{equation}
the coefficients $\alpha_p$ and $\beta_p$ are given by
\begin{equation}
\begin{split}
\alpha_p
=&\sqrt{\frac{\omega}{\omega^-}}
\frac{\Gamma(1-i\frac{\omega^-}{\Omega})\Gamma(-i\frac{\omega}{\Omega})}
{\Gamma(\frac{1}{2}-\frac{i}{2}(\frac{\omega^- +\omega}{\Omega}+c))\Gamma(\frac{1}{2}-\frac{i}{2}(\frac{\omega^- +\omega}{\Omega}-c))}, \\
\end{split}
\end{equation}
\begin{equation}
\begin{split}
\beta_p
=&\sqrt{\frac{\omega}{\omega^-}}
\frac{\Gamma(1-i\frac{\omega^-}{\Omega})\Gamma(i\frac{\omega}{\Omega})}
{\Gamma(\frac{1}{2}-\frac{i}{2}(\frac{\omega^- -\omega}{\Omega}+c))
\Gamma(\frac{1}{2}-\frac{i}{2}(\frac{\omega^- -\omega}{\Omega}-c))}. \\
\end{split}
\end{equation}
These coefficients satisfy $|\alpha_p|^2-|\beta_p|^2=1$.
In general, the coefficient $\beta_p$ does not become zero and the creation and annihilation operators at the in and out regions are mixed by the Bogoliubov transformation.
Thus, one finds that the in vacuum does not coincide with the out vacuum and particle creation occurs due to the external electric field.

\section{momentum spectrum of the produced particles \label{sec:Momentum_spectrum}}
Having introduced the new definition of asymptotic particles, we investigate particle creation in detail based on this particle picture.
We consider the situation in which the state is the in vacuum $|0\rangle_{\mathrm{in}}$ at the in region $t \rightarrow -\infty$.
Then, the expected number of produced particles with the momentum $p$ at the out region $t \rightarrow \infty$, is given by the vacuum expectation value of the operator $c_p^{\dagger} c_p$ about the state $|0\rangle_{\mathrm{in}}$,
\begin{equation}
\begin{split}
{}_{\mathrm{in}}\langle 0 | c_p^{\dagger} c_p |0 \rangle_{\mathrm{in}} =& |\beta_p|^2 (2\pi)^3 \delta^3 (\mathbf{p}=\mathbf{0}), \\
N_p= &\frac{(2\pi)^3}{\delta^3 (\mathbf{p}=\mathbf{0})} {}_{\mathrm{in}}\langle 0 | c_p^{\dagger} c_p |0 \rangle_{\mathrm{in}}, \\
=&|\beta_p|^2,
\end{split}
\end{equation}
where $V=\delta^3(\mathbf{p}=\mathbf{0})/(2\pi)^3$ is the total volume of space and $N_p$ represents the number of the produced particles per unit volume.
In our case, the specific expression for the distribution of the produced particles is given by
\begin{equation}
\begin{split}
|\beta_p|^2=&\frac{\sin \pi (\frac{1}{2}+\frac{i}{2}(\frac{\omega^- -\omega}{\Omega} +c)) \sin \pi (\frac{1}{2}+\frac{i}{2}(\frac{\omega^- -\omega}{\Omega}-c)) }
{\sinh \pi \frac{\omega^-}{\Omega} \sinh \pi \frac{\omega}{\Omega}}, \\
=&\frac{\sinh^2 \frac{\pi}{2}c+\cosh^2 \frac{\pi}{2} (\frac{\omega^- -\omega}{\Omega})}
{\sinh \pi \frac{\omega^-}{\Omega} \sinh \pi \frac{\omega}{\Omega}}.
\end{split}
\label{eq:distribution}
\end{equation}
From the expression for $\alpha_p$,
\begin{equation}
\begin{split}
|\alpha_p|^2=&\frac{\sin \pi(\frac{1}{2}+\frac{i}{2}(\frac{\omega^- +\omega}{\Omega}+c)) \sin \pi(\frac{1}{2}+\frac{i}{2}(\frac{\omega^- +\omega}{\Omega}-c)) }
{\sinh \pi \frac{\omega^-}{\Omega} \sinh \pi \frac{\omega}{\Omega}}, \\
=&\frac{\sinh^2 \frac{\pi}{2}c+\cosh^2 \frac{\pi}{2}(\frac{\omega^- +\omega}{\Omega})}{\sinh \pi \frac{\omega^-}{\Omega} \sinh \pi \frac{\omega}{\Omega}},
\end{split}
\end{equation}
we see that the coefficients satisfy the property of the Bogoliubov transformation $|\alpha_p|^2-|\beta_p|^2=1$.
This distribution coincides with that of preceding studies re-expressed by the kinetic momentum~\cite{Nikishov_pulse, Gitman7162, Mostepanenko_graphene}.
Furthermore, the same expression is found in Ref.~\cite{Fukushima}, in which the gauge condition is set to $A_3=0$ at $t \rightarrow \infty$.
This distribution is asymmetric about the momentum in the direction of the applied electric field. 
We now investigate the characteristic features of this distribution in detail.

In a time-dependent electric field, it is known that there are two mechanisms for particle creation.
One is where the virtual charged particles are accelerated by the electric field to the energy to become real particles, which is known as the Schwinger mechanism.
This process can be understood as some kind of tunneling process and it is a non-perturbative phenomenon.
Therefore, one cannot derive this phenomenon via the perturbative expansion of the classical external field $A_{\mu}(x)$.
The other mechanism is particle creation caused by the oscillation energy of the electric field, which is called dynamical pair creation.
This process is a multi-photon process in which the virtual charged particles gain energy via the scattering of external photons and a perturbative phenomenon.
In other words, particle creation occurs because of the effects of switching the electric field on and off.
Therefore, in this paper, we consider those situations in which these two mechanisms dominate each other, and investigate the parameter regions of the theory and characteristic features of the produced particles in each case.

First, we consider the case in which dynamical pair creation dominates.
In this case, because we consider the $\Omega$ as being large, the term
\begin{equation}
\frac{1}{\Omega}(\omega^- -\omega)=\frac{1}{\Omega} \biggl\{ \omega \Bigl[ 1+\frac{4 p_3}{\omega} \frac{eE}{\omega \Omega}+\Bigl(\frac{2eE}{\omega \Omega}\Bigr)^2 \Bigr]^{\frac{1}{2}} -\omega \biggr\}
\end{equation}
can be expanded in the following way, imposing the condition $\frac{eE}{\omega \Omega} \ll 1$,
\begin{equation}
\frac{1}{\Omega}(\omega^- -\omega) \simeq \biggl(\frac{2p_3}{\omega}+\frac{2eE}{\omega \Omega} \biggr) \frac{eE}{\Omega^2}.
\end{equation}
Thus, if $\frac{eE}{\Omega^2}\ll 1$, we can treat $\frac{1}{\Omega}(\omega^- -\omega)$ as a small quantity that is sufficiently smaller than unity.
This condition is satisfied in conjunction with the condition $\frac{eE}{\omega \Omega} \ll 1$, when we impose $\frac{\omega}{\Omega} \le \mathcal{O}(1)$.
Using this expansion, one can approximate the distribution $N_p$ as
\begin{equation}
\begin{split}
N_p \simeq & \frac{\bigl[\pi (\frac{eE}{\Omega^2})^2 \bigr]^2-\bigl[\frac{\pi}{2}\frac{(\omega^- -\omega)}{\Omega} \bigr]^2}
{\sinh^2 \pi \frac{\omega}{\Omega}}, \\
\simeq & \frac{ \bigl[\pi(\frac{eE}{\Omega^2})^2 \bigr]^2- \bigl[\pi (\frac{p_3}{\omega}+\frac{eE}{\omega \Omega} ) \frac{eE}{\Omega^2} \bigr]^2}
{\sinh^2 \pi \frac{\omega}{\Omega}}.
\end{split}
\end{equation}

\begin{figure}
 \includegraphics[width=9cm,clip]{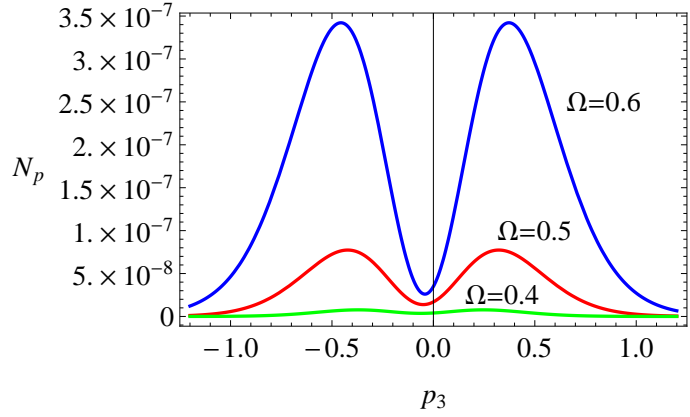}
 \caption{\label{fig:dynamical_pair_creation} Momentum spectrum of particles produced by dynamical pair creation with $\frac{eE}{m_{\perp}^2}=\frac{1}{40}$, and all in the units of $m_{\perp}$.
The number of produced particles becomes smaller as the frequency becomes smaller.
Thus, it establishes that particle creation is caused by the oscillation energy of the applied electric field.
}
 \end{figure}

In deriving the above approximate expression, we have imposed the following conditions:
\begin{equation}
\frac{eE}{\omega\Omega}\ll 1, \hspace{0.5cm} \frac{\omega}{\Omega} \le \mathcal{O}(1).
\label{eq:condition_dynamical}
\end{equation}
These conditions can be understood physically.
First, $\frac{eE}{\Omega}$ expresses the energy that a particle can obtain theoretically due to the acceleration by an electric field during the time-interval of the application of the electric field.
Then, the former condition expresses that the energy obtained from the electric field is larger than the threshold energy for becoming a real particle, and indicates that it is difficult for the Schwinger mechanism to occur.
On the other hand, the latter condition expresses that the oscillating energy of the electric field is not lower than the energy of the produced particle, and indicates that dynamical pair creation occurs easily.
Thus, these conditions (Eq.~(\ref{eq:condition_dynamical})) are those in which dynamical pair creation dominates the Schwinger mechanism.
In fact, the approximate expression for the distribution is given as a series of the small quantities $\frac{eE}{\omega\Omega}\ll 1$, which indicates that the particles are produced by the perturbative mechanism of dynamical pair creation.
The schematic picture of the distribution is shown in Fig.~\ref{fig:dynamical_pair_creation}.
The distribution is almost symmetric about the direction of the applied electric field.
This fact can be established from Eq.~(\ref{eq:distribution}), where the origin is located at $p_3=-\frac{eE}{\Omega}$, which is very small under the condition $\frac{eE}{\Omega^2} \ll 1$.

Next, we consider the case where the Schwinger mechanism dominates.
In this case, because we concentrate our attention on small $\Omega$, the hyperbolic function in Eq.~(\ref{eq:distribution}) will be approximated as the exponential function because of the large argument.
In fact, the denominator in Eq.~(\ref{eq:distribution}) can be approximated to the exponential function, if we impose the condition
\begin{equation}
\frac{m_{\perp}}{\Omega} \gg 1,
\end{equation}
where $m_{\perp}^2=m^2+p_{\perp}^2$.
For the numerator in Eq.~(\ref{eq:distribution}), we consider each case depending on the value of $p_3$.
Now that the distribution is symmetric about $p_3=-\frac{eE}{\Omega}$, in the following we consider only the case $p_3 \ge -\frac{eE}{\Omega}$.
This corresponds to $\omega^- > \omega$.

First, we consider the case where $p_3$ is in the region $-\frac{eE}{\Omega} \le p_3 < -m_{\perp}$.
Then, if we impose the condition
\begin{equation}
\frac{eE}{m_{\perp}\Omega} \gg 1,
\end{equation}
the expression
\begin{equation}
\frac{1}{\Omega}(\omega^- -\omega)=\frac{1}{\Omega} \biggl\{ \Bigl(p_3+\frac{2eE}{\Omega}\Bigr) \Bigl[ 1+\frac{m_{\perp}^2}{(p_3+\frac{2eE}{\Omega})^2} \Bigr]^{\frac{1}{2}}
-|p_3| \Bigl( 1+\frac{m_{\perp}^2}{p_3^2} \Bigr)^{\frac{1}{2}} \biggr\}
\end{equation}
can be expanded as
\begin{equation}
\frac{1}{\Omega}(\omega^- -\omega)=\frac{2eE}{\Omega^2}-\frac{2|p_3|}{\Omega}
-\frac{m_{\perp}^2}{\Omega}\frac{1}{|p_3| (\frac{2eE}{\Omega}-|p_3|)} \left(\frac{eE}{\Omega}-|p_3| \right).
\end{equation}
Since the contribution from the last two terms is always smaller than $-1$, we see that the dominant contribution in the numerator of the distribution Eq.~(\ref{eq:distribution}) comes from $\sinh^2 \frac{\pi}{2}c$.

Next, in the case where $p_3$ is in the region $-m_{\perp} \le p_3 \le |\mathcal{O}(m_{\perp})|$, the expression $\frac{1}{\Omega}(\omega^- -\omega)$ can be expanded in the following way, under the same condition $\frac{eE}{m_{\perp}\Omega} \gg 1$:
\begin{equation}
\begin{split}
\frac{\omega^- -\omega}{\Omega} \simeq &\frac{2eE}{\Omega^2}+\frac{p_3}{\Omega}+\frac{m_{\perp}^2}{4eE}-\frac{\omega}{\Omega}, \\
\simeq & \frac{2eE}{\Omega^2}+\frac{p_3}{\Omega}-\frac{\omega}{\Omega}, \\
=&\frac{2eE}{\Omega^2}-\left| \mathcal{O}(\frac{m_{\perp}}{\Omega})\right|,
\end{split}
\end{equation}
where in the second line, we utilized the fact that $\frac{m_{\perp}}{\Omega} \gg \frac{m_{\perp}^2}{eE}$, which comes from the condition $\frac{eE}{m_{\perp}\Omega} \gg 1$.
In a similar way, the last term is smaller than $-1$ and we find that the term $\sinh^2\frac{\pi}{2}c$ is dominant in the numerator in Eq.~(\ref{eq:distribution}) in this region of $p_3$.

We have imposed two conditions: $\frac{m_{\perp}}{\Omega} \gg 1$ and $\frac{eE}{m_{\perp}\Omega} \gg 1$.
From this condition, 
\begin{equation}
\frac{eE}{\Omega^2}=\frac{eE}{m_{\perp}\Omega}\frac{m_{\perp}}{\Omega} \gg 1
\end{equation}
is satisfied automatically.
Therefore, the term $\sinh^2 \frac{\pi}{2}c$ can be approximated as the exponential function.
Thus, in each region of $p_3$, the distribution can be expressed approximately as:
\begin{equation}
N_p = e^{-\pi (\omega^- +\omega-c)}.
\label{eq:distribution_simple}
\end{equation}
In the case where $p_3$ is in the region $p_3 > |\mathcal{O}(m_{\perp})|$, we find that the distribution $N_p$ is much smaller than that in the region $-\frac{eE}{\Omega} \le p_3 \le \mathcal{O}(m_{\perp})$.
Thus, we simply omit the case.

The distribution Eq.~(\ref{eq:distribution_simple}) can be expanded further in each region of $p_3$.
First, in the region $-\frac{eE}{\Omega} \le p_3 < -m_{\perp}$, $\frac{1}{\Omega}(\omega^- +\omega)-c$ is expanded as follows:
\begin{equation}
\begin{split}
\frac{\omega^- +\omega}{\Omega} -c \simeq & \frac{m_{\perp}^2}{2\Omega}\frac{1}{|p_3| (\frac{2eE}{\Omega}-|p_3|)}\frac{2eE}{\Omega}, \\
\simeq &\frac{m_{\perp}^2}{eE}\frac{1}{(1+\frac{\delta p_3 \Omega}{eE})(1-\frac{\delta p_3 \Omega}{eE})},
\end{split}
\end{equation}
where we have used $p_3=-\frac{eE}{\Omega}+\delta p_3$.
Furthermore, if $\frac{\delta p_3 \Omega}{eE}$ is smaller than unity,
\begin{equation}
\begin{split}
\frac{\omega^- +\omega}{\Omega} -c \simeq &\frac{m_{\perp}^2}{eE} \Bigl[ 1+\Bigl(\frac{\delta p_3 \Omega}{eE} \Bigr)^2 \Bigr].
\end{split}
\end{equation}
That is, the distribution is expressed as the Gaussian distribution function about $\delta p_3$.

Next, we consider the case where $p_3$ is in the region $-m_{\perp} \le p_3 \le |\mathcal{O}(m_{\perp})|$.
In this case, $\frac{1}{\Omega}(\omega^- +\omega)-c$ is expanded as
\begin{equation}
\begin{split}
\frac{\omega^- +\omega}{\Omega}-c =&\frac{m_{\perp}^2}{4eE}+\frac{p_3}{\Omega}+\frac{\omega}{\Omega}, \\
=&\frac{m_{\perp}^2}{4eE}+\frac{\mathcal{O}(m_{\perp})}{\Omega}.
\end{split}
\end{equation}
Since the condition $\frac{m_{\perp}^2}{eE} \ll \frac{m_{\perp}}{\Omega}$ holds, we find that the distribution is always suppressed exponentially relative to the peak value and is sufficiently small.

To summarize the above analysis, the distribution is given approximately by
\begin{equation}
\begin{split}
N_p =& e^{-\frac{\pi m_{\perp}^2}{eE}\frac{1}{(1+\frac{\delta p_3 \Omega}{eE})(1-\frac{\delta p_3 \Omega}{eE})} }, \\
\simeq & e^{-\frac{\pi m_{\perp}^2}{eE} [ 1+ (\frac{\delta p_3 \Omega}{eE} )^2 ] }.
\end{split}
\label{eq:distribution_approximation}
\end{equation}
It must be noted that as $p_{\perp}^2$ becomes larger, the condition $\frac{eE}{m_{\perp}\Omega} \gg 1$ is not satisfied.
However, in such a case, $N_p$ is already damped sufficiently and in what follows, we assume that the approximate expression Eq.~(\ref{eq:distribution_approximation}) is valid over all momenta.

\begin{figure}
 \includegraphics[width=8.3cm,clip]{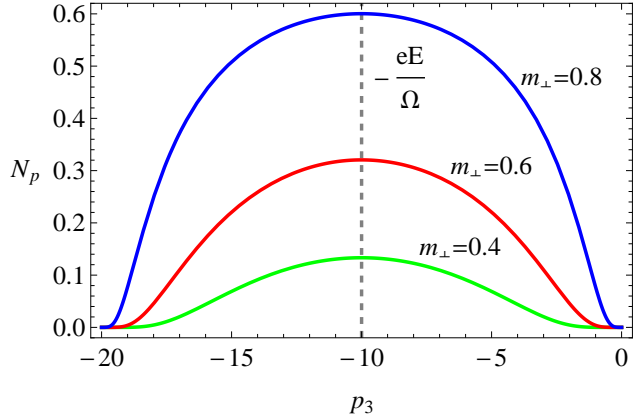}
 \caption{\label{fig:Schwinger_mechanism} Momentum spectrum of particles produced by the Schwinger mechanism with $\frac{e E}{\Omega^2}=100$ and all in the units of $\sqrt{e E}$.
As the strength of the field becomes stronger, the width of the distribution becomes broader, but damps exponentially as $p_3$ reaches both ends $p_3=-\frac{2eE}{\Omega}, 0$.
Thus, the produced particles are always distributed between $p_3=-\frac{2eE}{\Omega}$ and $p_3=0$.
}
 \end{figure}

Now, the approximate expression Eq.~(\ref{eq:distribution_approximation}) can be used to identify the width of the distribution.
Both expressions of Eq.~(\ref{eq:distribution_approximation}) indicate that the width will reach the order $\mathcal{O}(\frac{eE}{\Omega})$ for the first time when the magnitude of the electric field approaches the critical value $\frac{eE}{m_{\perp}^2} \sim 1$.
From the above consideration, we find the characteristic features of the distribution that $N_p$ is distributed between $-\frac{2eE}{\Omega} <p_3 <0$ with the peak amplitude given by $N_p=e^{-\frac{\pi m_{\perp}^2}{eE}}$, and that the width is determined by the strength of the external field $\frac{eE}{m_{\perp}^2}$.
The distribution becomes constant about $p_3$ with the width given by $\frac{2eE}{\Omega}$ when the electric field is extremely strong $\frac{eE}{m_{\perp}^2} \gg 1$.
These observations are verified numerically in Fig.~\ref{fig:Schwinger_mechanism}.

The parameter region where the Schwinger mechanism dominates has been given by
\begin{equation}
\frac{eE}{m_{\perp}\Omega} \gg 1, \hspace{0.5cm} \frac{m_{\perp}}{\Omega} \gg 1.
\end{equation}
These conditions are consistent with an intuitive picture in a similar way to dynamical pair creation.
That is, the former shows that the Schwinger mechanism occurs easily and the latter shows that dynamical pair creation is difficult to occur.
Note especially that the condition for the Schwinger mechanism to occur is not given by $\frac{eE}{\omega \Omega}$ but given by $\frac{eE}{m_{\perp}\Omega}$, which does not depend on $p_3$.
This may reflect the characteristic features of the particles produced by the Schwinger mechanism.
That is, in the Schwinger mechanism, particles are first always produced with zero momentum in the direction of the applied electric field $p_3=0$.
Then, these particles are accelerated in one direction, because of the applied electric field, to form the distribution between $-\frac{2eE}{\Omega} \le p_3 \le 0$.
In the strong electric field $\frac{eE}{m^2} \gg 1$, the spectrum becomes a nearly uniform distribution over $-\frac{2eE}{\Omega} \le p_3 \le 0$, which is qualitatively the same result as the T-constant field~\cite{Gitman7162}.

\section{Comparison with a T-constant electric field  \label{sec:Implication}}
The case with small $\Omega$ in our analysis, in which the Schwinger mechanism dominates, corresponds to particle creation in a constant electric field that is turned on and off adiabatically.
So it is interesting to compare results with those produced in a T-constant electric field~\cite{Gitman7162}.
In the preceding study, it was argued that a pulsed electric field produces the same particle spectrum as that in a constant field only after we take the limit of infinite-time interval~\cite{Nikishov_pulse, Nikishov_Nucl}.
However in our analysis, we find that particle spectrum by a pulsed electric field does not generally coincide with that by a constant field even when we take the infinite-time interval limit $\Omega \rightarrow 0$.
We claim that the disagreement comes from the misconception of canonical and kinetic momentums in the preceding study as follows.
First, if one uses the canonical momentum to express the distribution, as in the case of the preceding study, one obtains a symmetric distribution about the origin $p_3=0$.
When one takes the large-time interval here, the width of the distribution becomes increasingly broad and reaches a constant distribution at the infinite-time interval limit $T=\frac{1}{\Omega} \rightarrow \infty$.
Thus, one is subject to judge that the distribution $N_p$ becomes constant about $p_3$.
However, in our particle picture, in which the canonical momentum is identical to the kinetic ones, the origin of the distribution is shifted to $p_3=-\frac{eE}{\Omega}$.
In this case, when we take the limit of a large-time interval, the width becomes increasingly broad, but it is generally much narrower than $\mathcal{O}(\frac{eE}{\Omega})$, and at the same time the origin of the distribution is displaced further from $p_3=0$.
Thus, even if we take the limit $T \rightarrow \infty$, the distribution never becomes constant over all $p_3$.
Therefore, we argue that the result of the preceding study is an incorrect result, which is a consequence of the inappropriate manipulation of the limit $\frac{1}{\Omega} \rightarrow \infty$ with a canonical momentum fixed finite.

A T-constant field produces a constant momentum spectrum $N_p=e^{-\frac{\pi m_{\perp}^2}{eE}}$ over some finite momentum regions~\cite{Gitman7162}.
Since a pulsed electric field also produces almost constant spectrum for strong field regime $\frac{eE}{m^2} \gg 1$, these results will coincide.
In fact, in the strong field $\frac{eE}{m^2} \gg 1$, the distribution can be approximated as constant about $p_3$, $N_p=e^{-\frac{\pi m_{\perp}^2}{eE}}$ over the interval $-\frac{2eE}{\Omega} \le p_3 \le 0$.
Then, first the total number of produced particles $N$ is calculated as:
\begin{equation}
\begin{split}
N=&\int d^3 p N_p, \\
=&\int d^2 p_{\perp} \int_{-\frac{2eE}{\Omega}}^0 dp_3 N_p, \\
=&\frac{2 (eE)^2}{\Omega}e^{-\frac{\pi m^2}{eE}}.
\end{split}
\end{equation}
In a similar way, the vacuum-to-vacuum transition amplitude is calculated as:
\begin{equation}
\begin{split}
P_v=|{}_{\mathrm{out}} \langle 0 | 0 \rangle_{\mathrm{in}}|^2
=&e^{-\frac{V}{(2\pi)^3} \int d^3 p \ln(1+N_p)}, \\
=&e^{-\frac{V}{(2\pi)^3} \sum_{n=1}^{\infty} \frac{(-1)^{n+1}}{n} \int d^3 p N_p^n}, \\
\equiv &e^{-\int d^4 x w}, \\
\end{split}
\end{equation}
\begin{equation}
\int d^4 x w =\frac{V}{(2\pi)^3} \frac{2(eE)^2}{\Omega} \sum_{n=1}^{\infty} \frac{(-1)^{n+1}}{n^2}e^{-\frac{n\pi m^2}{eE}}.
\end{equation}
These results coincide with the bosonic version of the Schwinger's results, if we interpret the time interval of the applied electric field $T$ as $T=\frac{2}{\Omega}$~\cite{Schwinger}.

However, for a general strength of electric field, the Schwinger's results are no longer reproduced.
To see this, we first consider the case of the weak field $\frac{eE}{m^2} \ll 1$, where the Gaussian distribution approximation $N_p=e^{-\frac{\pi m_{\perp}^2}{eE} [ 1+(\frac{\delta p_3\Omega}{eE})^2 ]}$ is valid.
When $\delta p_3$ does not satisfy the condition $\frac{\delta p_3 \Omega}{eE} \ll 1$, this approximation is not valid.
However, under the condition of the weak field $\frac{eE}{m^2} \ll 1$, the distribution is damped exponentially before $\delta p_3$ reaches the inapplicable regions.
Therefore, in the weak field case, the Gaussian approximation is a good approximation over all momentum $p_3$.
In this case, the integral $\int d^3 p N_p^n$ is calculated as:
\begin{equation}
\begin{split}
\int d^3p N_p^n=&\int d^2 p_{\perp} dp_3 e^{-\frac{n \pi (m^2+p_{\perp}^2)}{eE} [1+(\frac{\delta p_3 \Omega}{eE})^2 ] }, \\
=&\frac{(eE)^2}{n\Omega} e^{-\frac{n\pi m^2}{eE}} \int_{-\infty}^{\infty} d(\delta \tilde{p}_3)  (\delta \tilde{p}_3^2+1)^{-1} e^{-\frac{n \pi m^2}{eE} \delta \tilde{p}_3^2},
\end{split}
\end{equation}
where we have defined the dimensionless variable $\delta \tilde{p}_3=\frac{\delta p_3 \Omega}{eE}$.
This expression is transformed to the complementary error function $\mathrm{erfc}(z)$, by (3.466, 1) in Ref.~\cite{Ryzhik}:
\begin{equation}
\int d^3 p N_p^n=\frac{(eE)^2}{n \Omega} \pi \mathrm{erfc}(\sqrt{\tfrac{n\pi m^2}{eE}}).
\end{equation}
Using the asymptotic expansion for the complementary error function $\mathrm{erfc}(x)=\frac{e^{-x^2}}{x \sqrt{\pi}}[1-\frac{1}{2x^2}+\dots]$, one can express the total number of produced particles and the vacuum-to-vacuum transition amplitude as follows:
\begin{equation}
\begin{split}
N=& \frac{(eE)^2}{\Omega} \pi \mathrm{erfc}(\sqrt{\tfrac{\pi m^2}{eE}}), \\
\simeq &\frac{(eE)^2}{\Omega} \left(\frac{eE}{m^2} \right)^{\frac{1}{2}} e^{-\frac{\pi m^2}{eE}}, 
\end{split}
\end{equation}
\begin{equation}
P_v= e^{-\int d^4 x w},
\end{equation}
\begin{equation}
\begin{split}
\int d^4 x w=& \frac{V}{(2\pi)^3} \sum_{n=1}^{\infty} \frac{(-1)^{n+1}}{n^2} \frac{(eE)^2}{\Omega}\pi \mathrm{erfc}(\sqrt{\tfrac{n\pi m^2}{eE}}), \\
\simeq & \frac{V}{(2\pi)^3}\frac{(eE)^2}{\Omega}\left( \frac{eE}{m^2} \right)^{\frac{1}{2}} \sum_{n=1}^{\infty} \frac{(-1)^{n+1}}{n^{\frac{5}{2}}} e^{-\frac{n\pi m^2}{eE}}.
\end{split}
\label{eq:summation}
\end{equation}
Thus, we find that the Schwinger's results are no longer reproduced in the weak field case $\frac{eE}{m^2} \ll 1$.
If we approximate the summation of Eq.~(\ref{eq:summation}) by the $n=1$ term due to the exponential suppression for $\frac{eE}{m^2} \ll 1$, both the total number of produced particles and the vacuum-to-vacuum transition amplitude are $(\frac{eE}{m^2})^{\frac{1}{2}}$ times smaller than the Schwinger's results.

If it is not the case $\frac{eE}{m^2} \ll 1$, then the Gaussian distribution approximation is no longer valid.
A more general expression is given by $N_p=e^{-\frac{\pi m_{\perp}^2}{eE} [(1+\frac{\delta p_3 \Omega}{eE})(1-\frac{\delta p_3 \Omega}{eE})]^{-1} }$.
This expression is damped exponentially as the momentum $p_3$ reaches each end $p_3 = -\frac{2eE}{\Omega}, \, 0$.
Therefore, we assume that the use of the above expression, delimited by the interval of $-\frac{2eE}{\Omega} \le p_3 \le 0$, is a good approximation.
Then, the integral $\int d^3 p N_p^n$ is calculated as:
\begin{equation}
\begin{split}
\int d^3 p N_p^n =& \int_{-\frac{2eE}{\Omega}}^{0} dp_3 \int d^2 p_{\perp} e^{-\frac{n\pi m_{\perp}^2}{eE} [ (1+\frac{\delta p_3 \Omega}{eE})(1-\frac{\delta p_3 \Omega}{eE}) ]^{-1} }, \\
=&\frac{(eE)^2}{n\Omega} 2 \int_{0}^{1} d(\delta \tilde{p}_3) [(1+\delta \tilde{p}_3)(1-\delta \tilde{p}_3)] e^{-\frac{n \pi m^2}{eE}\frac{1}{(1-\delta \tilde{p}_3)(1+\delta \tilde{p}_3)} }.
\end{split}
\end{equation}
The change of the variable $\frac{1}{(1+\delta \tilde{p}_3)(1-\delta \tilde{p}_3)}=t$ enables us to transform the expression further to
\begin{equation}
\int d^3 p N_p= \frac{(eE)^2}{n \Omega} \int_1^{\infty} dt t^{-\frac{5}{2}} (t-1)^{-\frac{1}{2}} e^{-\frac{n \pi m^2}{eE}t}.
\end{equation}
This integral can be converted to a Whittaker function $W_{\lambda, \, \mu}(z)$ by the formula (3.383, 4) in Ref.~\cite{Ryzhik},
\begin{equation}
\int d^3 N_p^n=\frac{\pi (eE)^2}{n\Omega} \left(\frac{n m^2}{eE} \right)^{\frac{1}{2}} e^{-\frac{n\pi m^2}{2eE}}W_{-1, 1}(\tfrac{n\pi m^2}{eE}).
\end{equation}
Thus, the total number of produced particles and the vacuum-to-vacuum transition amplitude are given, respectively, by:
\begin{equation}
N=\frac{\pi (eE)^2}{\Omega} \left(\frac{m^2}{eE}\right)^{\frac{1}{2}} e^{-\frac{\pi m^2}{2eE}} W_{-1,1}(\tfrac{\pi m^2}{eE}),
\label{eq:total_number}
\end{equation}
\begin{equation}
P_v= e^{-\int d^4 x w},
\end{equation}
\begin{equation}
\int d^4 x w =\frac{V}{(2\pi)^3}\frac{\pi(eE)^2}{\Omega} \left(\frac{m^2}{eE}\right)^{\frac{1}{2}} \sum_{n=1}^{\infty}\frac{(-1)^{n+1}}{n^{\frac{3}{2}}} W_{-1,1}(\tfrac{n \pi m^2}{eE}) e^{-\frac{n \pi m^2}{2eE}}.
\label{eq:vacuum_transition}
\end{equation}

Now, let us check the validity of these expressions.
First, in the weak field case $\frac{eE}{m^2} \ll 1$, using the asymptotic expansion for a Whittaker function with small $\frac{eE}{m^2}$, $W_{\lambda, \, \mu}(z) \sim e^{-\frac{z}{2}} z^{\lambda} (1+\frac{\mu^2-(\lambda-\frac{1}{2})^2}{z}+\dots)$, we find
\begin{equation}
\int d^3p N_p^n \simeq \frac{(eE)^2}{n^{\frac{3}{2}} \Omega} \left(\frac{eE}{m^2}\right)^{\frac{1}{2}} e^{-\frac{n\pi m^2}{eE}}.
\label{eq:Schwinger_smaller}
\end{equation}
Thus, the expressions for $N$ and $\int d^4 x w$ coincide exactly with those of the Gaussian approximation.
Next, in the strong field case $\frac{eE}{m^2} \gg 1$, the asymptotic form $W_{\lambda, \, \mu}(z) \sim \frac{\Gamma(2\mu)}{\Gamma(\frac{1}{2}+\mu-\lambda)} z^{-\mu+\frac{1}{2}} e^{-\frac{z}{2}}$, $[ \, \mu >0 \,]$ can be used to calculate the total number of produced particles as:
\begin{equation}
N=\frac{4}{3} \frac{(eE)^2}{\Omega} e^{-\frac{\pi m^2}{eE}}.
\end{equation}
This expression agrees with the Schwinger's result by the interpretation of the time interval $T=\frac{1}{\Omega} \mathcal{O}(1)$.
Thus, the expression $N_p=e^{-\frac{\pi m_{\perp}^2}{eE} [1-\delta \tilde{p}_3^2]^{-1} }$ is a good approximation over all values of $\frac{eE}{m^2}$.

An analysis using the Gaussian distribution approximation has already been made~\cite{Gitman7162}.
There, it was concluded that in the strong field regime results by a pulsed electric field do not coincide with those by a T-constant electric field.   
However, we argue that this conclusion is an incorrect consequence of the fact that they used the Gaussian approximation beyond the limit of its applicability $\frac{eE}{m^2} \ll 1$.
In fact, we use the more general expression $N_p=e^{-\frac{\pi m_{\perp}^2}{eE} [1-\delta \tilde{p}_3^2]^{-1} }$, and demonstrate that the pulsed field reproduces the result of a T-constant field in the strong field regime $\frac{eE}{m^2} \gg 1$ even at a finite-time interval.

Our analysis shows that pulsed and T-constant fields produce qualitatively the same results in strong field but they differ in general field strength.
The reason of the difference can be understood as a sensitivity of momentum spectra against electric field strength.
Generally, an electric field with strength $E$ produces a momentum spectrum $N_p=e^{-\frac{\pi m_{\perp}^2}{eE}}$ by the Schwinger mechanism.
Now, let us consider a weak field case $E/n$ with $n>1$.
This electric field produces a distribution $N_p=\bigl( e^{-\frac{\pi m_{\perp}^2}{eE}} \bigr)^n$.
In the strong field regime, the value in the parenthesis nears $1$, therefore power of $n$ does not change the value so greatly.
However in the weak field regime, the value in the parenthesis is already very smaller than $1$, and power of $n$ further make the value very small.
Thus, the difference in results between pulsed fields and T-constant fields is understood as the sensitivity of the spectrum against the field strength.
We tend to interpret the applied time interval $T$ as $1/\Omega$ given a pulsed electric field.
However in general field strength, the spectrum is so sensitive against the variation of the field strength during the time interval $1/\Omega$ that we cannot interpret the applied time interval as $1/\Omega$.
In fact in the preceding study, the time interval of the pulsed field is interpreted as $T=\frac{1}{\Omega} (\frac{eE}{m^2})^{\frac{1}{2}}$ in the weak field regime $\frac{eE}{m^2} \ll 1$, to conform with the Schwinger's formula~\cite{Gitman7162}.

We can understand the difference in results between pulsed electric fields and T-constant fields physically as whether nonperturbative effects of high-frequency photons of external fields exist or not.
In fact, the Fourier transform of the pulsed and T-constant fields, $E(t)=\int \frac{d \omega}{2\pi} \widetilde{E}(\omega)e^{i \omega t}$, are respectively given by
\begin{equation}
\widetilde{E}_p(\omega)=\frac{E\pi \omega}{\Omega^2 \sinh \pi \omega/2\Omega},  \;\; \; \; \widetilde{E}_c(\omega)=\frac{2E}{\omega} \sin\omega T/2.
\end{equation}
That is, the T-constant field consists of photons with all the momentum scale, on the other hand, the pulsed field consists of photons with definite momentum scale ranging from 0 to $\Omega$.
Thus, we could argue that the constant distribution over some momentum range produced by T-constant fields is formed by an influence of nonperturbative effects of high-frequency photons in external fields.
However, the effects become invisible in the strong field regime.

\section{Conclusions \label{sec:Conclusion}}
In this paper, we elaborated on particle creation from a vacuum by a single pulse of an electric field in scalar quantum electrodynamics.
We first define the asymptotic particles in which the canonical momentum is identical to the kinetic ones.
Based on this particle picture, we identified the parameter regions of the theory in which dynamical pair creation and the Schwinger mechanism dominate, respectively.

Furthermore, we derive analytical expressions for the various characteristics of particle creation, such as the vacuum-to-vacuum transition amplitude where the Schwinger mechanism dominates.
Then, we compare our results with those produced by a T-constant electric field.
In the preceding study, it was argued that a pulsed electric field produces the same particle spectrum as those in a constant field only after we take the limit of the infinite-time interval~\cite{Nikishov_pulse, Nikishov_Nucl}.
However, our detailed analysis reveals that the results of pulsed fields and T-constant fields coincide in the strong field regime even at a finite-time interval $1/\Omega$.
In general field strength the results differ due to the sensitivity of the Schwinger mechanism against external field strength.
We interpreted the difference as a nonperturbative effect of high-frequency photons in external fields, though the effect becomes invisible in the strong field regime.
Thus in the actual experiments by strong laser techniques, it will be important to take into account the duration that the maximum field strength is applied and high-frequency photon effects, in addition to field strength itself in observing the produced particles.


\bibliography{basename of .bib file}

\begin{thebibliography}{99}
\bibitem{Schwinger}
   J. Schwinger, Phys. Rev. {\bf 82}, 664 (1951).



\bibitem{Nikishov}
   A. I. Nikishov, Zh. Eksp. Teor. Fiz. {\bf 57}, 1210 (1969), [Sov. Phys. JETP {\bf 30}, 660 (1970)].
\bibitem{Nikishov_pulse}
   N. B. Narozhny and A. I. Nikishov, Yad. Fiz. 11, 1072-1077 (1970), [Sov. J. Nucl. Phys. 11, 596 (1970)].
\bibitem{Itzykson}
   E. Brezin and C. Itzykson, Phys. Rev. D {\bf 2}, 1191 (1970).
\bibitem{Grib}
   A. A. Grib, V. M. Mostepanenko and V. M. Frolov, Teor. Mat. Fiz. {\bf 13}, 1207 (1972), [Theor. Math. Phys. {\bf 13}, 1207 (1972)].
\bibitem{Gitman}
   V. G. Bagrov, D. M. Gitman and Sh. M. Shwartsman, Zh. Eksp. Teor. Fiz. {\bf 68}, 392 (1975), [Sov. Phys. JETP {\bf 41}, 191 (1975)].




\bibitem{Nikishov2}
   N. B. Narozhnyi and A. I. Nikishov, Zh. Eksp. Teor. Fiz. {\bf 65}, 862 (1973), [Sov. Phys. JETP {\bf 38}, 427 (1974)].
\bibitem{Mostepanenko}
   V. M. Mostepanenko and V. M. Frolov, Yad. Fiz. {\bf 19}, 885 (1974), [Sov. J. Nucl. Phys. {\bf 19}, 451 (1974)].




\bibitem{Dunne}
   R. Schutzhold, H. Gies and G. Dunne, Phys. Rev. Lett. {\bf 101}, 130404 (2008), [arXiv:0807.0754~[hep-th]].
\bibitem{Alkofer}
   M. Orthaber, F. Hebenstreit and R. Alkofer, Phys. Lett. B {\bf 698}, 80 (2011), [arXiv:1102.2182~[hep-ph]].
\bibitem{Schutzhold}
   C. Fey and R. Schutzhold, Phys. Rev. D {\bf 85}, 025004 (2012), [arXiv:1110.5499 [hep-th]].




\bibitem{Allor}
   D. Allor, T. D. Cohen and D. A. McGady, Phys. Rev. D {\bf 78}, 096009 (2008), [arXiv:0708.1471~[cond-mat.mes-hall]].
\bibitem{Beneventano}
   C. G. Beneventano, P. Giacconi, E. M. Santangelo and R. Soldati, J. Phys. A {\bf 42}, 275401 (2009), [arXiv:0901.0396 [hep-th]].
\bibitem{Mostepanenko_graphene}
   G. L. Klimchitskaya and V. M. Mostepanenko, Phys. Rev. D {\bf 87}, 125011 (2013), [arXiv:1350.5700 [hep-th]].


\bibitem{Nikishov_Nucl}
   A. I. Nikishov, Nucl. Phys. {\bf B21}, 346 (1970).






\bibitem{Gitman7162}
   S. P. Gavrilov and D. M. Gitman, Phys. Rev. D {\bf 53}, 7162 (1996), [hep-th/9603152].


\bibitem{Fukushima}
   K. Fukushima, F. Gelis and T. Lappi, Nucl. Phys. A {\bf 831}, 184 (2009), [arXiv:0907.4793 [hep-ph]].



\bibitem{Ryzhik}
   I. S. Gradshteyn and I. M. Ryzhik, \textit{Table of Integrals, Series, and Products, Fourth Edition} (Academic, New York, 1965).





\end{thebibliography}

\end{document}